\documentclass[12pt]{article}
\usepackage{mathtext}
\usepackage{graphicx}
\usepackage[T2A]{fontenc}
\usepackage[utf8]{inputenc}
\usepackage{amsfonts}
\usepackage{amssymb}
\usepackage{amsmath}
\usepackage{pgfplots}
\usepackage{braket}
\usepackage{tikz}
\usepackage{float}
\usepackage{multicol}
\newcommand{\g}{\gamma}

\begin{document}
    \title{Numerical analysis of the influence of initial and external conditions on the association of artificial monovalent atoms}
    
    \author{Chen Ran$^{1}$, Yuri Ozhigov$^{1,2}$ \\
    {\it 
    1.Moscow State University of M.V.Lomonosov, VMK Faculty, Russia} 
    \\
    {\it 2. Institute of Physics and Technology RAS (FTIAN), Moscow}
    \\
    }
\maketitle

\begin{abstract}
The chemical dynamics scene is the most important application area of computer simulation. We show that electrons jump between potential wells of different depths (new molecular orbitals hybridized by atomic orbitals with different energies) under the influence of temperature (phonons) and photonic phenomena. Overcome exponentially growing computational complexity. In our paper, we experimented with state space selection algorithms.

\end{abstract}

\section{Introduction}

Quantum Chemistry Computing The complexity of computation has grown exponentially and has long been a difficult problem for computational mathematics. Various effects of temperature, electromagnetic field and decoherence complicate the process and calculation of the reaction, increasing the computational load, for example, polaron rotation (for example, see\cite{Polariton1}), photon blocking and surrounding optical cavities in Jaynes. - The role of finite-dimensional Cummings models (\cite{Polariton2} and \cite{Polariton3}) The article considers the Jaynes-Cummings model for electrons jumping between atoms (potential wells of different depths).

The focus of our work is more related to collective effects and complex chemical scenarios than to detailed fine microscopy studies (e.g. in Polariton4, within the Holstein-Tavis-Cummings model, to quantum considerations of polyatomic systems from a point of view).

First, we assume that in the QED cavity there is a pair of hydrogen atoms forming a molecule, and one electron jumps between the potential well formed by the two atoms. Therefore, we use the Jaynes-Cummings model to solve the Schrödinger equation(\cite{Polariton5}, \cite{Polariton6}, \cite{Polariton7}):

\begin{equation}
i\hbar|\dot{\Psi}\rangle=H|\Psi\rangle
\end{equation}

In the JC model, the Hamiltonian H is expressed as:

\begin{equation}
H_{JC}=\hbar\omega a^+a+\hbar\omega \sigma^+\sigma+g(\sigma^+ +\sigma )(a+ a^+)
\end{equation}

where $a,a^+$ are the field operators of photon annihilation and creation, and $\sigma,\sigma^+$ are the relaxation and excitation operators of the atom. In the case of weak interactions in chemical reactions, $g/\omega\ll 10^{-2}$ is approximated, so the RWA approximation can be used for the Hamiltonian H:

\begin{equation}
H_{JC}^{RWA}=\hbar\omega a^+a+\hbar\omega \sigma^+\sigma+g(\sigma^+a+\sigma a^+),
\label{RWA}
\end{equation}

We also use the Lindblad equation to calculate open quantum systems:

\begin{equation}
i\hbar\dot{\rho}=[H,\rho]+i{\cal L}(\rho),{\cal L}(\rho)=\sum\limits_{i=1}^{N-1}\g_i(A_i\rho A_i^+-\frac{1}{2}\{A_i^+A_i\rho,\rho A_i^+A_i\})
\end{equation}
The loss of coherence is expressed as the decoherence coefficient $A_i$ (\cite{BP}).

\section{Frequencies of electron transition into potential wells of different depths.}
This time we simulate the case of an electron (different depths of potential wells) in a covalent bond formed by an oxygen atom and a hydrogen atom. We want to observe electrons tunneling between two atoms.

	\begin{figure}[H]
		\centering
		\includegraphics[width=1.5in]{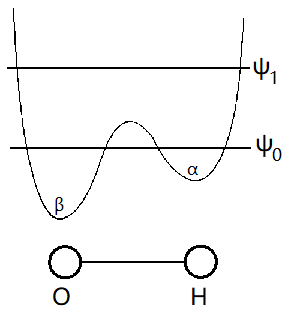}
		\caption{ground state and excited states}
		\label{psi0-1OH.png} 
	\end{figure} 

Here is the potential energy:$\alpha>\beta$,that is
$|\Psi_0\rangle = \frac{\alpha|O\rangle+\beta|H\rangle}{\sqrt{\alpha^2+\beta^2}}$,
$|\Psi_1\rangle = \frac{-\beta|O\rangle+\alpha|H\rangle}{\sqrt{\alpha^2+\beta^2}}$    

where $|O\rangle$ and $|H\rangle$ are the excited state orbitals of oxygen and hydrogen atoms, respectively.

\subsection{Hamiltonian and operators}

$a^+$ -photon creation operator, $a$-photon annihilation operator.

$\sigma^+$ -atomic excitation operator, $\sigma$ – atomic relaxation operator

$H=\hbar\omega a_{\omega}^+a_{\omega}+\hbar\omega\sigma_{\omega}^+\sigma_{\omega}+g_{mol}(a_{\omega}^+\sigma_{\omega}+a_{\omega}\sigma_{\omega}^+)$

\subsection{Computational Mathematical Analysis Solution}

We use this formula to calculate how the quantum state changes over time under the influence of the Hamiltonian H(\cite{Tav}):

$|\Psi(t)\rangle=\sum_j e^{\frac{-i}{\hbar}E_jt}\lambda_j|\psi_j\rangle$,где $|\Psi(0)\rangle=|0\rangle_{ph}|O\rangle=\frac{\alpha|0\rangle_{ph}|\Psi_0\rangle-\beta|0\rangle_{ph}|\Psi_1\rangle}{\sqrt{\alpha^2+\beta^2}}$

So we get:

$|\Psi(t)\rangle=\frac{2\alpha^2+\beta^2(e^{iE_1t}+e^{iE_2t})}{2(\alpha^2+\beta^2)}|0\rangle|O\rangle+\frac{\alpha\beta(e^{iE_1t}-e^{iE_2t})}{2(\alpha^2+\beta^2)}|1\rangle|O\rangle+\frac{2\alpha\beta-\alpha\beta(e^{iE_1t}+e^{iE_2t})}{2(\alpha^2+\beta^2)}|0\rangle|H\rangle+\frac{-\beta^2(e^{iE_1t}-e^{iE_2t})}{2(\alpha^2+\beta^2)}|1\rangle|H\rangle$

	\begin{figure}[H]
		\centering
		\includegraphics[width=5.5in]{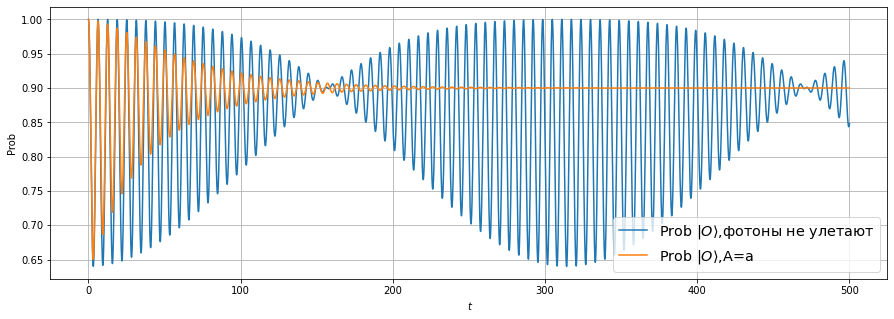}
		\caption{The probability of an electron appearing in a potential well near an oxygen atom $P(|O\rangle)$}
		\label{fig:OH1e-1.png} 
	\end{figure} 

where $P(|O\rangle)$(there are photon escape and decoherence A=a) tends to $y=\alpha^2$. We see that electrons have a high probability of ending up in deeper potential wells.

\subsection{A pair of electrons tunnels in a double potential well formed by an oxygen atom and a hydrogen atom}
The Hamiltonian has the following form:

$H=\hbar\omega a_{\omega\uparrow}^+a_{\omega\uparrow}+\hbar\omega a_{\omega\downarrow}^+a_{\omega\downarrow}+H_{\sigma}+H_{\omega}$

$H_{\sigma}=\hbar\omega\sigma_{\omega_1\uparrow}^+\sigma_{\omega_1\uparrow}+\hbar\omega\sigma_{\omega_1\downarrow}^+\sigma_{\omega_1\downarrow}+\hbar\omega\sigma_{\omega_2\uparrow}^+\sigma_{\omega_2\uparrow}+\hbar\omega\sigma_{\omega_2\downarrow}^+\sigma_{\omega_2\downarrow}$

$H_{\omega}=g_{mol}(a_{\omega\uparrow}^+(\sigma_{\omega_1\uparrow}+\sigma_{\omega_2\uparrow})+a_{\omega\uparrow}(\sigma_{\omega_1\uparrow}^++\sigma_{\omega_2\uparrow}^+)+a_{\omega\downarrow}^+(\sigma_{\omega_1\downarrow}+\sigma_{\omega_2\downarrow})+a_{\omega\downarrow}(\sigma_{\omega_1\downarrow}^++\sigma_{\omega_2\downarrow}^+))$

For convenience of calculations, we will assume that the directions of the spins of two electrons are opposite and do not change the direction of the spin.

The initial state is
$|\Psi_{start}\rangle=|0\rangle_{ph\uparrow}|0\rangle_{ph\downarrow}|O{\uparrow}\rangle|H{\downarrow}\rangle$.

\begin{figure}[htp]
\centering
    \begin{minipage}[t]{0.50\textwidth}
        \centering
        \includegraphics[width=1\textwidth]{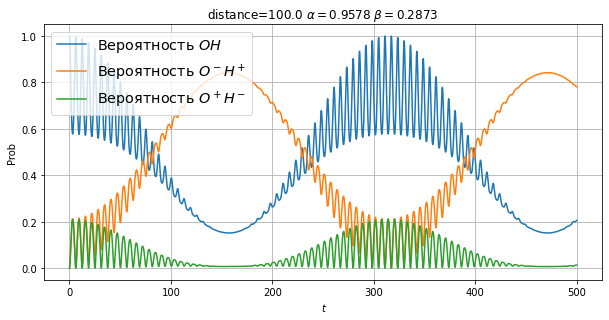}
        \caption{Probability of 2 electrons appearing, no decoherence}
        \label{OH2e-1.png}
    \end{minipage}
    \begin{minipage}[t]{0.49\textwidth}
        \centering
        \includegraphics[width=0.8\textwidth]{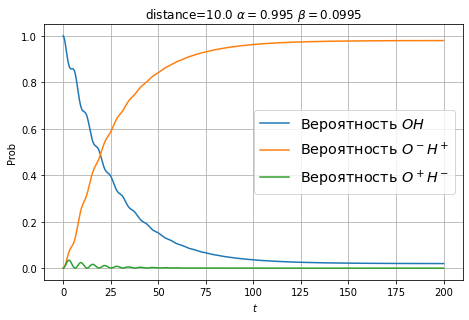}
        \caption{Add decoherence factor A=a}
        \label{OH2e-2.png}
        \end{minipage}
\end{figure}

Here we see that electron tunneling does change periodically, and the amplitude is related to the potential energies $\alpha$ and $\beta$. Since $\alpha > \beta$, the amplitude of $P(O^-H^+)$ is greater than that of $P(O^+H^-)$. The average amplitude of $P(OH)$ is the largest (one electron is in the potential well of the oxygen atom and the other electron is in the potential well of the hydrogen atom), which means that the oxygen and hydrogen atoms remain electrically neutral. The larger the $P(OH)$, the stronger the bond.

In the case of photon escape (see\cite{Oz3}), if $\alpha\gg\beta$, when time $t\rightarrow\infty$ we have:
$P(O^-H^+)=\alpha^4 \gg P(OH)=2\alpha^2\beta^2 \gg P(O^+H^-)=\beta^4$

\section{Modeling the effect of temperature on the formation of bonds in liquids using phonons}

\begin{figure}[htp]
\centering
    \begin{minipage}[t]{0.46\textwidth}
        \centering
        \includegraphics[width=0.8\textwidth]{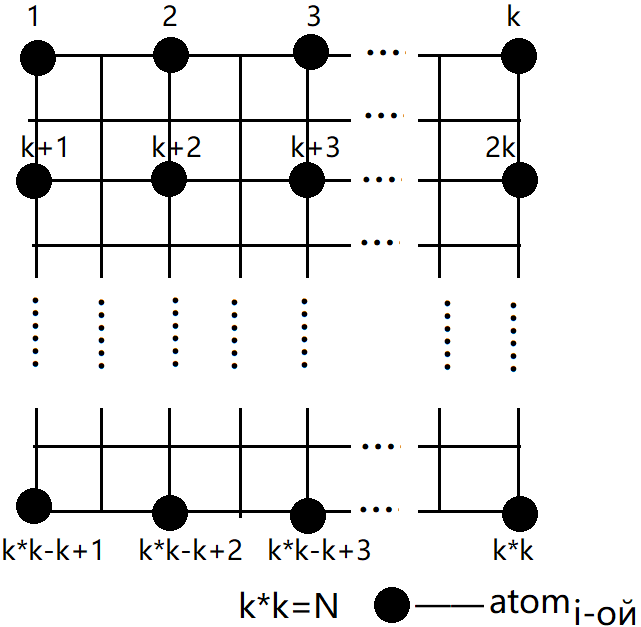}
        \caption{Atoms are arranged in a grid}
        \label{grid.png}
    \end{minipage}
    \begin{minipage}[t]{0.46\textwidth}
        \centering
        \raisebox{0.3\height}{\includegraphics[width=0.4\textwidth]{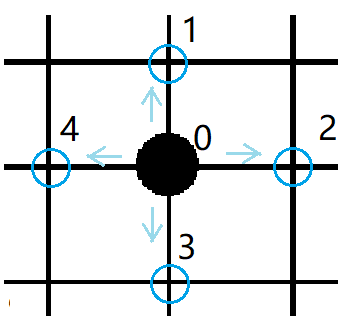}}
        \caption{4 directions}
        \label{move.png}
        \end{minipage}
\end{figure}

In this part we take into account the influence of phonons (temperature, kinetic energy, etc.) on coupling. Let's assume that the atoms are neatly arranged on a grid and can only move over a small range (for example, the observed object is a solid or a liquid). There are k rows * k columns = N atoms.

We assume that each atom can only move in four directions, up, down, left, and right, when the target position of motion is not occupied by other atoms (a Coulomb repulsive force will occur between nuclei when the distance is too close).

\subsection{Operators and Hamiltonians}

$b$:Phonon operator

Hamiltonians:
$H = H_{tun}+H_{cov}+\hbar\omega b^+b+\hbar\omega\sigma_{cov}^+\sigma_{cov}$

$H_{tun(at_i)} =g_{tun}\sum_{j=1}^4(\sigma_{0\rightarrow j}+\sigma_{j\rightarrow0})check_{cov(ati)}$

$H_{cov}=g_{cov}\sum^{sum_{cov}}_i{b^+\sigma_{cov_i}}+b\sigma_{cov_i}^+$

Description of each state:$|b\rangle_{phono}|pos_1\rangle|pos_2\rangle…|pos_N\rangle|cov_1\rangle…|cov_M\rangle$

\subsection{Numerical calculation results}

\begin{figure}[H]
		\centering
		\includegraphics[width=5in]{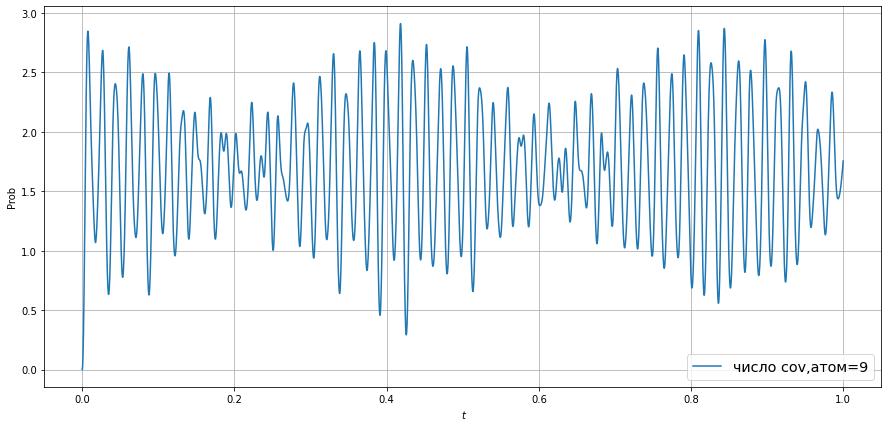}
		\caption{Expectations for covalent bond formation change over time}
		\label{9atgcov=10.png} 
	\end{figure} 

We see that the image changes periodically, approximately symmetrically around y=1.75. Let us denote this axis of symmetry as $y=k$. And in the case of a certain initially given number of phonons, no matter how much the coefficient $g_{cov}$ of the Hamiltonian is increased, this axis $y=k$ remains almost unchanged, and only slightly changes the oscillation of the frequency curve. Therefore, we can know that the range of variation of the expected value is related only to temperature (number of phonons).

\section{We will consider both photon and phonon effects on chemical bonds.}

Let's consider the case when water molecules form hydrogen bonds. As shown in the figure below, the “-OH” portion of one water molecule combines with an oxygen atom in another water molecule to form a hydrogen bond. Atoms can change their distance from each other by moving.

Each atom has an electron involved in bonding. Thus, 3 electrons are involved in the calculation. The two oxygen atoms are labeled "$O_1$" and "$O_2$" respectively. The orbital of the first oxygen atom hybridizes with the orbital of the hydrogen atom to form two energy levels $\Psi_1^1$ and $\Psi_0^1$, the orbital of the second oxygen atom hybridizes with the orbital of the hydrogen atom to form $\Psi_1^2$ and $\Psi_0^2 $- two energy levels.
Based on this, we use the following Hamiltonian for the calculation:

\begin{figure}[H]
		\centering
		\includegraphics[width=2.5in]{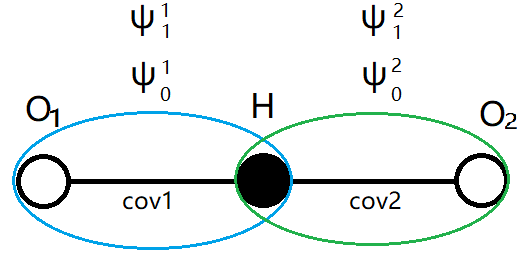}
		\caption{The -OH group forms a hydrogen bond with another oxygen atom}
		\label{oho.png} 
	\end{figure}

The hybridization orbitals of each pair $|\Psi_1^1\rangle$ and $|\Psi_0^1\rangle$ are orthogonal to each other, which is expressed by the following formula:

$|\Psi_0^i\rangle = \frac{\alpha|O_i\rangle+\beta|H\rangle}{\sqrt{\alpha^2+\beta^2}}$,
$|\Psi_1^i\rangle = \frac{-\beta|O_i\rangle+\alpha|H\rangle}{\sqrt{\alpha^2+\beta^2}},i=1,2$

\subsection{Operators and Hamiltonians}

$\sigma_{mol_i}$:Operators that change electron energy levels

$\sigma_{mol_i}^+|\Psi_0^i\rangle=|\Psi_1^i\rangle,\sigma_{mol_i}|\Psi_1^i\rangle=|\Psi_0^i\rangle,i=1,2$

$\sigma_{tun_i}$:Tunnel operator of the i-th atom. The distance between atoms can mutually change, and two atoms can form a chemical bond only when they are in close proximity.

$\sigma_{spin}$:Electron spin operators.

Hamiltonian:

$H = H_{tun}+H_{cov}+H_{spin}+H_{mol}+\hbar\omega b^+b+\hbar\omega a_{mol}^+a_{mol}+\hbar\omega a_{spin}^+a_{spin}+\Sigma{\hbar\omega \sigma_{cov_i}^+ \sigma_{cov_i}}$

The initial state:

$|\Psi(0)\rangle=|2\rangle_{a_{mol}}|1\rangle_{a_{spin}}|1\rangle_{b}|\Psi_0^1\uparrow\rangle_{O_1}|\Psi_1\uparrow\rangle_{H}|\Psi_0^2\downarrow\rangle_{O_2}|not,close\rangle_{cov_1}|exist,close\rangle_{cov_1}$

\subsection{Numerical calculation results}

\begin{figure}[H]
		\centering
		\includegraphics[width=5.8in]{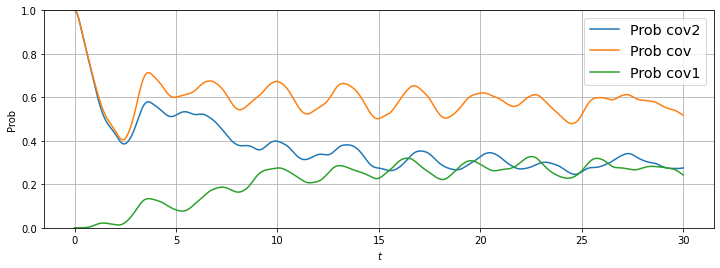}
		\caption{Graph of the probability of bond formation versus time}
		\label{oho-11.png} 
	\end{figure} 
 
\section{State space selection algorithm}
When the number of states is large, the calculations become very heavy. Therefore, we can select the portion of states with larger amplitude in the space of all states to model the calculus.

\begin{figure}[H]
		\centering
		\includegraphics[width=4in]{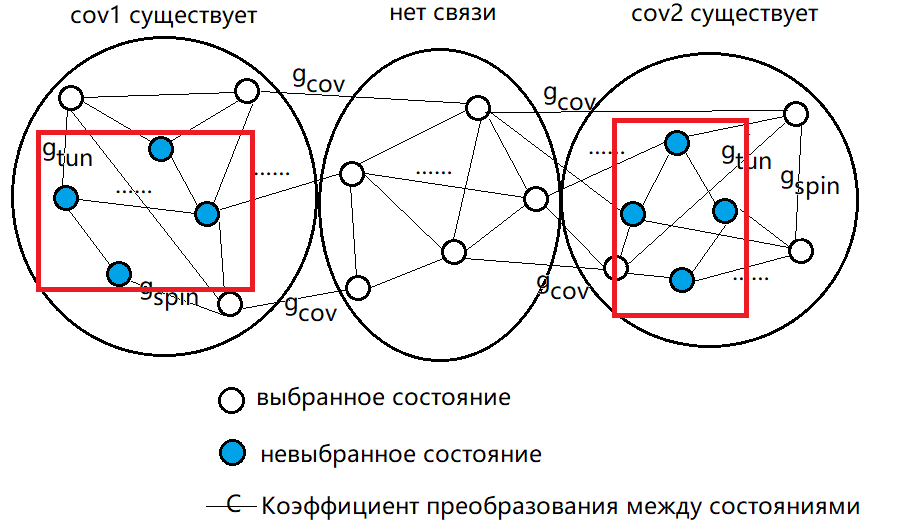}
		\caption{Some states with larger amplitude}
		\label{stateselection.png} 
	\end{figure} 

Each circle represents a state. The lines between the states mean that they can be directly converted into each other, and the numbers on the lines are the coefficients (Hamiltonian quantities). From Fig. \ref{oho.png} we can easily divide all states into three categories: presence of the first bond, presence of the second bond and no bond (only three electrons). However, during the selection process, some states have small amplitudes, but have a large influence on the selected state (the “not cov” part in the picture below). If we select a state only according to amplitude strength and store only the transformation relation between the selected states, we get an "unrelated" plot:

Therefore, we chose the optimization method:

Step 1: Check the graph connectivity every time

Step 2: If the graph is not connected, add all "neighbor states" $|\Psi_{neigh}\rangle$ of the selected state to the set and return to step 1 for looping; If the graph is connected, then end the loop and use these selected states for calculations .

$|\Psi_{neigh}\rangle$ of some state $|\Psi\rangle$: a state that can directly transform with this state through the Hamiltonian H, that is, $H|\Psi\rangle = \sum_i{\lambda_i|\ Psi_{neigh_i}\rangle}$

\begin{figure}[htp]
\centering
    \begin{minipage}[t]{0.46\textwidth}
        \centering
        \includegraphics[width=1\textwidth]{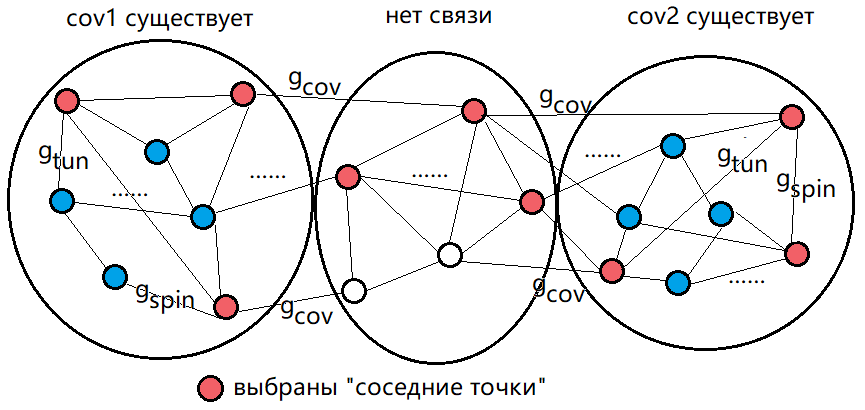}
        \caption{Neighboring points selected}
        \label{neighbor_points.png}
    \end{minipage}
    \begin{minipage}[t]{0.46\textwidth}
        \centering
        \raisebox{0.1\height}{\includegraphics[width=1\textwidth]{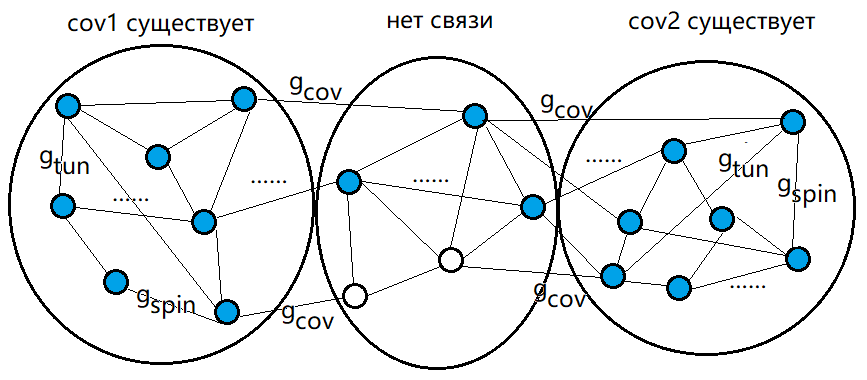}}
        \caption{New connected graph}
        \label{a_connected_graph.png}
        \end{minipage}
\end{figure}

\subsection{Numerical calculation results}

\begin{figure}[H]
		\centering
		\includegraphics[width=5.5in]{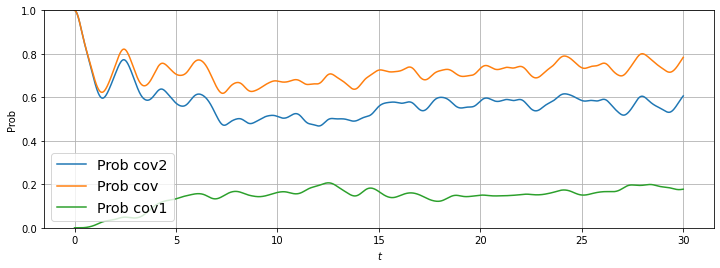}
		\caption{Graph of expected number of bonds formed versus time (using reduced state space)}
		\label{new2oho.png} 
	\end{figure} 

After comparison, we found that similar curve trends can be obtained using reduced state space, which is very useful for studying evolution between states. But the disadvantage is that using a new state space will change the position of the final stable fluctuation of the curve. This shows that the state space selection algorithm still needs improvement.

\section{Conclusion}
From a quantum point of view, we observe the phenomenon of electrons jumping between atoms in a hybrid orbital under the influence of light and temperature. Electron hopping, in turn, affects the formation of chemical bonds and the appearance of ions. We use phonons to participate in the formation of chemical bonds. Due to the influence of the depth of the potential well, the probability of electrons appearing near oxygen atoms is greater, which explains why hydrogen ions $H^+$ are more easily formed. We then show the transformation of the hydroxyl position due to electron hopping in the case of mutual hybridization of orbitals between the hydroxyl group and another oxygen atom. This shows that when a large number of water molecules are arranged in a regular manner (for example, in a crystalline state such as ice), the molecular structure can change flexibly. Finally, we used a state space selection algorithm. The pictures in the article are the result of choosing 1/5 the size of the original state space, which significantly reduces the computational load and can significantly improve efficiency. Because computational complexity grows exponentially as the state space increases. There is still room for further improvement of the algorithm. The current version can perfectly restore the shape and trend of the curve, but due to state reduction, the final stable value of the curve will be changed, and is expected to be improved in the future.

\end{document}